# An Assessment of Organics Detection and Characterization on the Surface of Europa with Infrared Spectroscopy

Ishan Mishra[1,2], Nikole Lewis[1], Jonathan Lunine[2,3], and Kevin P. Hand[2]
[1] Department of Astronomy and Carl Sagan Institute, Cornell University, 122 Sciences Drive, Ithaca, NY 14853, USA; ishan.mishra@jpl.nasa.gov
[2] Jet Propulsion Laboratory, California Institute of Technology, Pasadena, CA 91109, USA
[3] Division of Geological and Planetary Sciences, California Institute of Technology, Pasadena, CA 91125, USA

## Abstract

Organics, if they do exist on Europa, may only be present in trace amounts on the surface. NASA's upcoming mission Europa Clipper is going to provide global, high-quality data of the surface of Europa in the near-infrared (NIR), specifically the 3–5 $\mu$m region, where organics are rich in spectroscopic features. In this work we investigate Europa Clipper's ability to constrain the abundance of selected trace species of interest that span different chemical bonds found in organics, such as C–H, C=C, C≡C, C=O, and C≡N, via NIR spectroscopy in the 3–5 $\mu$m wavelength region. We simulate reflectance spectra of these trace species mixed with water ice, at varying signal-to-noise ratio and abundance fractions. The evidence for the trace species in a mixture is evaluated using two approaches: (1) calculating average strength of absorption feature(s), and (2) Bayesian model comparison (BMC) analysis. Our simulations show that sharp and strong spectroscopic features of trace (∼5% abundance by number) organic species should be detectable at >3$\sigma$ significance in Europa Clipper–quality data. A BMC analysis pushes the 3$\sigma$ detection threshold of trace species even lower to <1% abundance. We also consider an example with all trace species mixed together, with overlapping features, and BMC is able to retrieve strong evidence for all of them and also provide constraints on their abundance. These results are promising for Europa Clipper's capability to detect trace organic species, which would allow correlations to be drawn between the composition and geological regions with possibly endogenic material.

*Unified Astronomy Thesaurus concepts:* Planetary science (1255); Remote sensing (2191); Infrared spectroscopy (2285); Europa (2189); Bayesian statistics (1900); Surface composition (2115)

## 1. Introduction

Constraining the presence and abundance of organics on the surface of Europa is critical for determining the habitability of its subsurface ocean (S. A. Kattenhorn & L. M. Prockter 2014), as well as a primary objective of upcoming missions like Europa Clipper (D. L. Blaney et al. 2017; S. M. Howell & R. T. Pappalardo 2020). Underneath its long-lived, salty, liquid-water ocean, it's likely that there is water–rock interaction, providing an environment where chemical disequilibrium can be exploited as an energy source for potential life (K. P. Hand et al. 2009; J. I. Lunine 2017). However, one aspect of assessing Europa's habitability, which has been unconstrained so far, is the presence and distribution of organics (here defined as molecules with C–H, C=C, C≡C, C=O, and C≡N bonds) on this icy moon's surface and its subsurface ocean.

While we await missions that can go sample Europa's surface and ocean directly (K. P. Hand & C. R. German 2018; K. P. Hand et al. 2022), remote sensing is our best tool to try to constrain Europa's organic content and distribution. Water ice, as well as heavily hydrated non-ice material that dominates some areas of Europa (T. B. McCord et al. 1998; R. W. Carlson et al. 2009), has low reflectance at longer wavelengths in the near-infrared (NIR; ≈3–5 $\mu$m). This makes NIR a key wavelength regime for detecting trace abundances of organics because strong organic absorptions in this region can still produce detectable features even on a low-reflectance continuum. The 3–5 $\mu$m region is particularly diagnostic for functional groups such as C–H, C=O, and O–H that are common in organic molecules. Other Europa Clipper instruments focus on different scientific goals: the UVS instrument targets far-ultraviolet emissions from neutral and ionized sulfur, oxygen, and hydrogen in Europa's surrounding environment, while E-THEMIS will acquire thermal infrared images (7–70 $\mu$m) to constrain the thermophysical properties of the surface but does not provide spectroscopic measurements (S. M. Howell & R. T. Pappalardo 2020). Therefore, the Mapping and Imaging Spectrometer (MISE) on Europa Clipper provides the only direct spectroscopic capability to identify surface organics via their diagnostic absorption features.

Although organics have not been detected yet on the surface of Europa, an aliphatic hydrocarbon CH stretching band has been reported on the other Galilean satellite, Callisto, and a 4.57 $\mu$m feature, attributed to C≡N, is found in spectra of both Ganymede and Callisto (T. B. McCord et al. 1997). Two possible features (at 2.05 and 2.17 $\mu$m) were also found in the NIMS spectra (J. B. Dalton et al. 2003) that could be due to N–H combination bands of an amide, but these features are close to the noise level and difficult to differentiate from other materials, such as salts. In another discovery of relevance to Europa, Cassini found low-mass organic compounds in the Enceladean ice grains: nitrogen-bearing, oxygen-bearing, and aromatic (N. Khawaja et al. 2019). While the evidence for plumes on Europa remains tentative (W. B. Sparks et al. 2016; P. M. Schenk 2020), the possibility of fresh material on the surface of Europa via other regions of extrusions, such as chaos terrains (S. K. Trumbo et al. 2019), provides upcoming missions with compelling avenues to constrain the organic composition of the subsurface ocean.

Organics, if they do exist on Europa, may only be present in trace amounts on the surface. They would likely be embedded in a water-ice matrix or in other hydrated non-ice materials, as some regions of Europa, such as recently emplaced terrains and parts of the trailing hemisphere, are nearly ice-free and rich in spectral features of Europa's other non-ice components like hydrated sulfates, chlorides, and oxides, such as $CO_2$, $SO_2$, and $H_2O_2$ (R. W. Carlson et al. 2009). A rough upper limit on the abundance of $CH_2$ groups, $[CH_2]/[H_2O] < 1.5 \times 10^{-3}$, was estimated from their non-detection in Galileo/NIMS spectra of Europa (K. P. Hand et al. 2009). However, this constraint is optimistic and should be interpreted as a generous upper bound under ideal assumptions, especially given the low signal-to-noise ratio (SNR) of NIMS data in the 3–5 $\mu$m region. The number of $CH_2$ groups expected from meteoritic infall and burial by gardening is estimated to be a factor of three lower than this upper limit, or about 500 ppm. K. P. Hand et al. (2009) also state that if hydrocarbons are found locally on Europa with abundances greatly exceeding tenth-of-a-percent levels, they may point to an oceanic, rather than exogenic, source of organic material. Abundances, relative to water, of complex organics in the ice grains in Enceladus's plume (∼2%; N. Khawaja et al. 2019) and simple volatiles like HCN and $CH_3OH$ in comets (∼0.01%–10%; M. J. Mumma & S. B. Charnley 2011) can also inform our expectations for the abundances of organics that might be present on Europa's surface.

As for the effect of irradiation on organics on Europa's surface, an electron irradiation experiment of hydrocarbons mixed with water by K. P. Hand & R. W. Carlson (2012) showed that the resultant mixture was dominated by single carbon bonds, covalent bonds with hydrogen, bonds with –OH (hydroxyl), bonds with oxygen (C–O), or double bonds with oxygen (carbonyl). These results suggest that unsaturated bonds and functional groups associated with small volatile hydrocarbons (e.g., alkenes) are quickly destroyed upon irradiation, while longer-chain aliphatics with C–H, C–O, and C=O functional groups are more resilient and can persist as irradiation residues. Of course, these findings are based on simplified laboratory conditions and cannot fully replicate the complexities of Europa's surface environment, but they highlight which classes of organic molecules are more likely to survive and be detectable.

Given both the technical challenges of the Galileo mission and the harsh radiation environment of Europa, it is perhaps not surprising that organics have not yet been detected (R. W. Carlson et al. 2009). However, observations in the NIR (at spatial resolutions of approximately hundreds of kilometers per pixel) have started with JWST (S. K. Trumbo & M. E. Brown 2023; S. Trumbo et al. 2023; G. L. Villanueva et al. 2023), and high spatial resolution observations (less than a kilometer per pixel) are on the horizon within the next decade from the Europa Clipper (S. M. Howell & R. T. Pappalardo 2020) and Jupiter Icy Moons Explorer (JUICE) missions (O. Grasset et al. 2013). Specifically for Europa Clipper, mapping the composition of organics on Europa's surface is a primary objective (D. L. Blaney et al. 2017). Moreover, the high spatial resolution that MISE has to offer (<10 km per pixel for global mapping, approximately tens of meters at closest approach; D. L. Blaney et al. 2017) will allow correlations to be drawn between the composition and geological features like linea and chaos terrains, or even plume deposits, where endogenic material from the subsurface ocean is likely present (R. W. Carlson et al. 2009; L. C. Quick & M. M. Hedman 2020).

In this work, we present the first detection feasibility study of trace organic species on the surface of Europa, primarily as a function of the SNR of the spectroscopic data, with a specific focus on Europa Clipper's MISE instrument. Not only do the analysis tools presented in this work provide a glimpse of what is achievable with the high-quality spectroscopic data that Europa Clipper will collect, but also these tools can be used to set requirements for spectroscopic observations of Europa from other facilities, to ensure robust characterization of its surface composition. We investigate detectability of organics through analysis of simulated reflectance data of mixtures of water ice, the dominant background species on Europa, and various trace species. Our selected trace species are shown in Table 1 and span different chemical bonds of interest found in organics, such as C–H, C=C, C≡C, C=O, and C≡N, and henceforth collectively will be referred to as "trace organic species." Figure 1 highlights the difficulty of detecting trace species even in wavelength regions where they are spectroscopically rich. The 3–5 $\mu$m is rich in features, even for the simple molecules included in this work. However, in the presence of a dominant background species (here, water ice), the features of the trace species are weak, on the order of a percent or less in reflectance units. The challenge here is to pick out the weak signals of a particular molecule, in the presence of other major and minor candidates, and also estimate their abundance. Even a moderate grain size of 50 $\mu$m provides sharp features and strong detections for most of the trace organics considered in this work, especially when employing Bayesian model comparison (BMC). Reproducing the detection analysis presented here for varying grain sizes, or for

**Table 1**
The Central Wavelengths of Absorption Features of Trace Organic Species Used in This Work, as Estimated from Spectra Simulated Using Optical Constants Available at Temperatures Close to the Europan Ambient Temperature of ∼130 K (see Section 2.1)

| Species | Absorption Feature Location(s) ($\mu$m) |
|---|---|
| $C_2H_2$ | **4.25**, **4.85** |
| $C_2H_4$ | 3.26, 3.36, 3.76, **4.11**, 4.43, 4.62, **4.90** |
| $C_2H_6$ | 3.28, 3.35, 3.38, 3.64, **3.78**, **3.91**, **4.26**, 4.41, 4.52 |
| $C_3H_6O$ | 3.33, 3.36, 3.41, 3.59, 3.87, 3.99, **4.07**, **4.42**, **4.70**, 4.77 |
| $CH_3OH$ | **3.94**, **4.06**, 4.29, 4.45, 4.58, 4.70, **4.91** |
| HCN | **4.26**, **4.42**, **4.74**, 4.84 |
| $NH_3$ | **4.55**, **4.88** |
| $C_2N_2$ | 3.87, **4.26**, **4.52**, **4.62**, 4.77, 4.88 |

**Note.** Features present near the water-ice Fresnel peak around 3.1 $\mu$m and the volume scattering peak near 3.5 $\mu$m—specifically around 3.16 and 3.33 $\mu$m—are not shown here, as this region is consistently dark and makes organic features difficult to resolve owing to overlapping ice behavior. The prominent features that were used in calculating the $\sigma_{fs}$ metric for detection (see Section 2.2) are in bold. These were chosen by an automated Python routine that identified and ranked absorption features in a given species' mixed spectrum with water ice, based on their prominence, and were further confirmed through visual inspection.

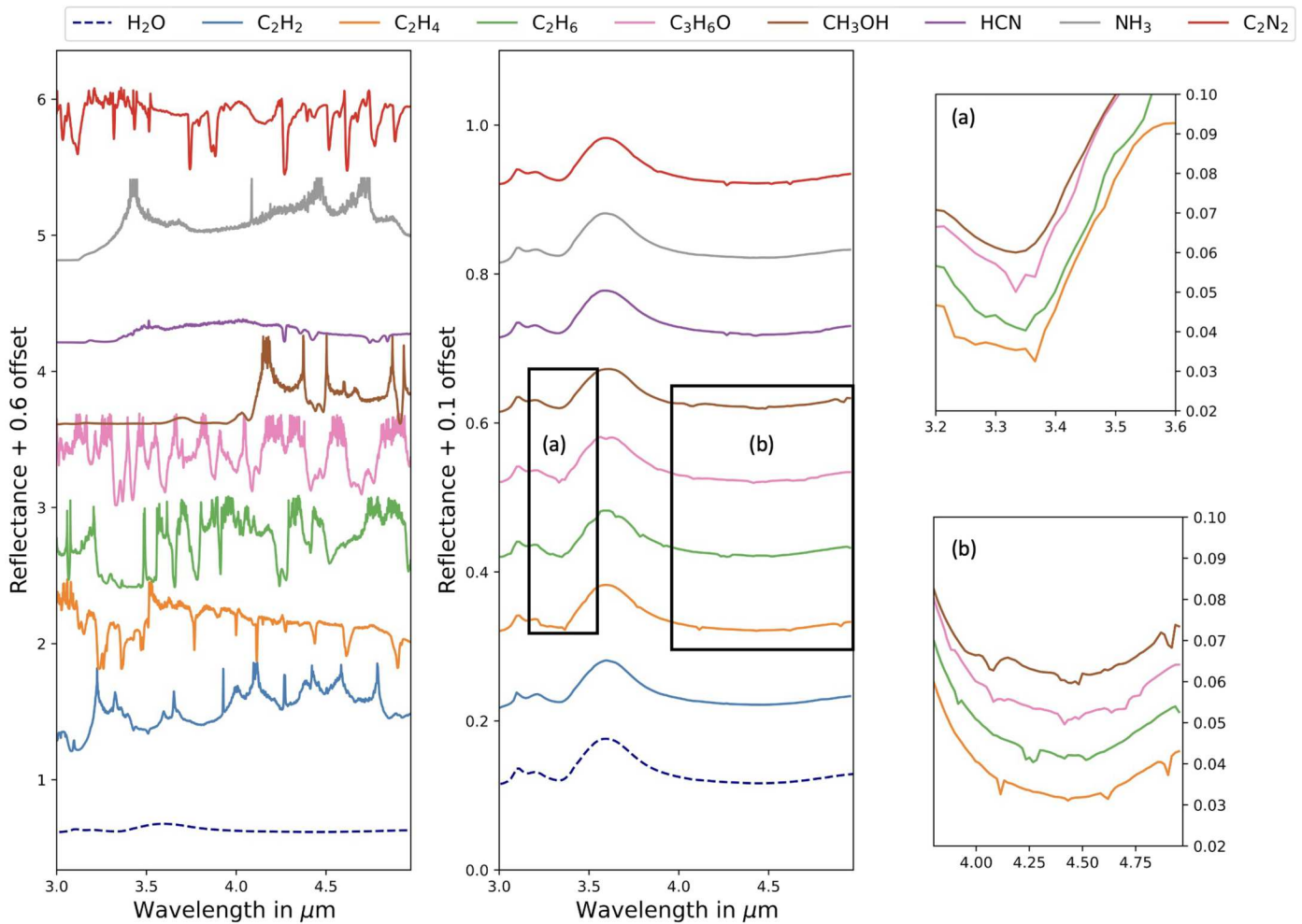

**Figure 1.** Left panel: simulated, noise-free spectra of water ice (dashed dark-blue curve) and selected trace species of interest. Middle panel: simulated spectra of non-water-ice species mixed with water ice, at an abundance fraction of 5%. The dashed dark-blue curve is a pure water-ice spectrum for reference. Right panel: zoom-in of the (a) 3.2–3.6 $\mu$m and (b) 3.8–5.0 $\mu$m regions of four spectra from the middle panel. While organics are spectrally rich in the 3–5 $\mu$m wavelength range, on Europa we expect them to be present in trace amounts in a background of water ice, which leads to weak absorption signatures (on the order of a few percent in strength) on top of a mostly water-ice-rich continuum. Other heavily hydrated, non-ice materials may also be plausible background components, and exploring their effects on detectability is an interesting direction for future work. The key physical parameters used in these simulations are as follows: incidence angle = emergence angle = 45°; phase angle = 90°; average grain size = 50 $\mu$m. Reflectance is expressed in units of radiance factor.

complex grain size distributions, could further refine the detectability limits of these species, but this is beyond the scope of the present work.

In Section 2, we describe our reflectance model (Hapke bidirectional reflectance model) and discuss two approaches—calculating the strength of absorption feature(s) versus BMC—for inferring the presence/absence of a trace species in a reflectance spectrum. We also present a new diagnostic tool in Section 2.4, called a "detectability heatmap," that allows us to visualize the detection significance of a trace species as a function of abundance and SNR and to compare different approaches of detection. In Sections 3.1 and 3.2 we investigate these heatmaps for all our selected trace organic species. Next, in Section 3.3, we discuss a complex example with many trace species mixed together with water ice and present results for a BMC analysis of the complex spectrum. Finally, in Section 3.4, we explore the sensitivity of sharp organic features to physical parameters like observation geometry angles, grain size of regolith, and porosity of regolith, followed by conclusions in Section 4.

## 2. Methodology

To investigate the detectability of trace organic species signatures in near-infrared reflectance spectra, we simulate observations of a planetary regolith with water ice and one or more trace species mixed together, and the model is described in Section 2.1. We choose water ice as the dominant background species in our simulations, as it is the dominant component in regions on Europa not dominated by particle irradiation products such as sulfuric acids and salts (I. Mishra et al. 2021b). While some recent terrains on Europa, as observed by Galileo SSI, appear rich in "dark" material that is likely non-ice and possibly more abundant than water ice in those areas, this material is believed to be highly hydrated and spectrally similar to water ice in the 3–5 $\mu$m region. Therefore, the choice of background species should not significantly affect our conclusions, as hydrated salts and acids believed to be present on Europa's surface are spectrally similar to water ice beyond 3 $\mu$m and lack the sharp features characteristic of organics (e.g., S. De Angelis et al. 2020, 2021). It should also be noted that surface grain size is influenced by a combination of processes, including particle irradiation and micrometeoroid bombardment, the latter of which causes mechanical gardening and sputtering that tend to produce finer-grained material, particularly on the leading hemisphere. As such, detectability of surface species is expected to vary across Europa's surface. The effect of grain size is briefly discussed in Section 3.4, but a more detailed analysis of detectability as a function of geography and surface processing is beyond the scope of this work.

To evaluate the evidence of a particular trace species in simulated data, we use two approaches, measuring the prominence (or amplitude with respect to the noise level) of their strongest absorption features (described in Section 2.2) and performing a BMC analysis of models with and without the species of interest (described in Section 2.3). In Section 2.4, we present a visual tool, called the "detectability heatmap," that helps us understand the detectability of a trace species within the SNR limits of any given instrument.

### 2.1. Simulating Reflectance Spectra with the Hapke Model

The standard Hapke equation (B. Hapke 1981; I. Mishra et al. 2021b, 2021a) provides an analytical approximation to the bidirectional reflectance of a planetary regolith:

$$r_F = K \frac{\omega(\lambda)}{4} \frac{\mu_0}{(\mu + \mu_0)}$$
$$\times [P(g, \lambda) + H(\omega, \mu/K) H(\omega, \mu_0/K) - 1], \quad (1)$$

where $r_F$ is the radiance factor, which is the ratio of the brightness of a surface to that of a perfectly diffuse surface (Lambertian) illuminated and observed at a 0° incidence angle, $\mu$ is the cosine of the emergence angle $e$, $\mu_0$ is the cosine of the incidence angle $i$, $g$ is the phase angle, $K$ is the porosity coefficient, $\omega$ is the single-scattering albedo, $P$ is the particle phase function, and $H$ is the Ambartsumian–Chandrasekhar function that accounts for the multiply scattered component of the reflection (S. Chandrasekhar 1960).

For most of the simulations presented here, we have fixed the observation geometry parameters, incidence, emission, and phase angles, to moderate values of 45°, 45°, and 90° respectively (in Section 3.4 we investigate the sensitivity to these parameters). Just like in the analysis of Galileo spectra of Europa presented in I. Mishra et al. (2021b), here we ignore parameters related to backscattering and other opposition effects, as they become important only in the regime of very small phase angles for the coherent backscatter effect and typically less than a few tens of degrees for the shadow-hiding opposition effect (B. Hapke 2012b, 2021). In addition, we ignore the photometric contribution of macroscopic roughness, which Hapke characterizes with a mean topographic slope angle, $\bar{\theta}$. The global average values of $\bar{\theta}$ derived from Voyager and Galileo imaging data vary little from 16° to 22° among different works (A. J. Verbiscer et al. 2013). The general effect of topography is to decrease the amount of reflected light coming toward the observer, and this effect gets more pronounced with increasing phase angle and increasing roughness of the surface. At the moderate photometric geometry used in this work ($i = 45°$, $e = 45°$, $g = 90°$), the effects should be small enough to be ignored. For the phase function $P$, we use a two-parameter Henyey–Greenstein function (L. G. Henyey & J. L. Greenstein 1941), which has been shown to be representative of a wide variety of planetary regoliths, including the icy particles on Europa (B. Hapke 2012a). Since we are interested in specific behavior associated with changes in species abundance, we fix the porosity coefficient $K$ to be 1. However, the effects of the observation angles and porosity on a reflectance spectrum are briefly discussed in Section 3.4.

The most important Hapke parameter of interest in this work is the single-scattering albedo $\omega$, which governs the wavelength-dependent contribution of a species in the total reflectance of a mixture of species. The equations for the single-scattering albedo come from the equivalent-slab approximation (B. Hapke 2012a) and are dependent on the optical constants (the real and imaginary refractive indices) of each species in the model, along with their average grain size (fixed at 50 $\mu$m for all simulations in this work; for more discussion on grain size see Section 3.4). The selection of the species used in this work was crucially governed by the availability of their optical constants, measured at temperatures close to the Europan temperature range of 80–130 K (J. R. Spencer et al. 1999) and at our wavelengths of interest (∼3−5 $\mu$m). For water ice, we use a modified version (Roger Clark, personal communication) of the optical constants of crystalline water ice published by R. M. Mastrapa et al. (2009). R. N. Clark et al. (2012) found that the optical constants of R. M. Mastrapa et al. (2009) contain an underestimation in the baseline transmission data in the 3.2–5 $\mu$m wavelength region. The choice between amorphous and crystalline ice should not matter for the purpose of this study, as they are spectroscopically similar in the 3–5 $\mu$m wavelength range.

The optical constants for the trace organic species, listed in Table 1, come from the online repository[4] of the Cosmic Ice Laboratory at NASA GSFC and have been measured at cryogenic temperatures of ∼60–120 K. The $CO_2$ optical constants, at 179 K, are from E. Quirico & B. Schmitt (1997) and E. Quirico et al. (1999), and the $SO_2$ optical constants, at 125 K, are from B. Schmitt et al. (1994, 1998). The $CO_2$ and $SO_2$ optical constants are available in the Solid SSHADE (Spectroscopy Hosting Architecture of Databases and Expertise) repository[5] (B. Schmitt et al. 2018).

For a detailed description of all the parameters in the Hapke equation used here (Equation (1)), including the equations used to calculate $\omega$ for a mixture of species, please see the Appendix of I. Mishra et al. (2021a).

To convert the high spectral resolution and noise-free spectrum produced by the Hapke model into realistic observations, we bin the spectrum down to the resolution of a given spectrometer and add Gaussian noise to the model for a given SNR. Figure 2 presents an example of such simulated reflectance data, where a model reflectance spectrum of a simple mixture of water and HCN (hydrogen cyanide) to match both the spectral resolution and sampling of MISE (10 nm) and Gaussian noise is added to it, corresponding to an SNR of 50, which is within the range of expected SNR of MISE (D. L. Blaney et al. 2017). Three prominent HCN features at 4.26, 4.42, and 4.74 $\mu$m are clearly visible in the simulated data, suggesting that MISE's spectral resolution and expected SNR are sufficient to resolve weak but sharp features such as those in Figure 2.

---

[4] https://science.gsfc.nasa.gov/691/cosmicice/constants.html
[5] https://www.sshade.eu/

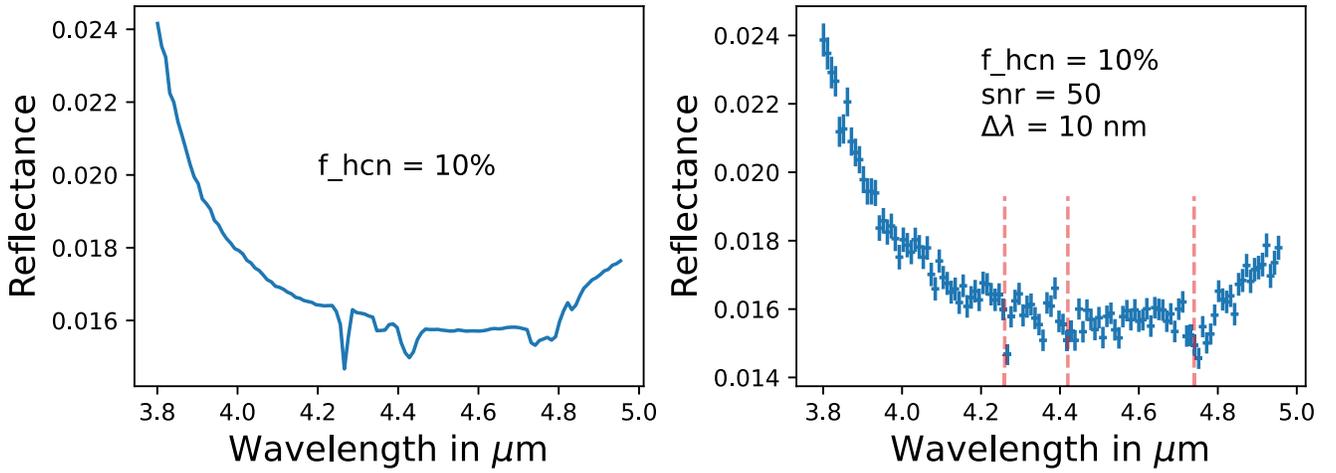

**Figure 2.** Left panel: a reflectance spectrum model of water ice and HCN mixture is shown, with the abundance (by number) fraction of HCN (f_hcn) = 10%. Right panel: spectroscopic data simulated using the model shown in the left panel, by adding Gaussian noise to it at SNR = 50 and binning the model down to the 10 nm spectral resolution ($\Delta\lambda$) of MISE. The dashed vertical red lines indicate the locations of prominent HCN features that are also clearly visible in this MISE quality data. The physical parameters used in the model are as follows: incidence angle = emergence angle = 45°; phase angle = 90°; average grain size = 50 $\mu$m. Reflectance units is radiance factor.

### 2.2. Evaluating Evidence for a Trace Species via Prominence of Its Spectroscopic Features

We first evaluate the evidence for the presence of a trace species via a simple, "by-eye" method. For a given species, we look for an absorption or a trough in the spectral curve at the expected wavelengths of the species' feature(s). For example, if evaluating evidence for HCN, we will look for "dips" at 4.26 and 4.42 $\mu$m, the expected locations of the two strongest absorption features of HCN in the 3–5 $\mu$m wavelength range (Figure 2).

For any given spectrum, we can calculate the amplitude of features (band depth) at the absorption wavelength(s) $\lambda_i$ of our species of interest (we calculate band depth using the `prominences` routine in Python's `scipy` package, which measures how much a peak (or a trough) stands out from the surrounding baseline signal). Using the values of these absorption features, we then calculate their average prominence or "strength":

$$\sigma_{\rm fs} = \frac{1}{N}\sum_{i=1}^{N} \frac{\text{amplitude of absorption feature at wavelength } \lambda_i}{\text{noise at wavelength } \lambda_i}, \quad (2)$$

where $N$ is the number of features/wavelengths we are looking for, "noise at wavelength $\lambda_i$" is the average noise of the simulated data points within the width of the feature, and "fs" means "feature strength," since we are evaluating the average strength of absorptions at the expected features of our species of interest. Here noise at a given data point is calculated by dividing its reflectance by the chosen SNR. For example, $\sigma_{\rm fs} = 3$ would mean that at their expected wavelengths there are absorption features whose amplitude is around 3 times the noise level (aka 3$\sigma$ strength). While the symbol $\sigma$ is typically used for standard deviation, here we use it to denote the strength of the features to keep it consistent with the Bayesian inference terminology (discussed in the following section), where $\sigma$ is used as a metric for confidence of detection.

### 2.3. Evaluating Evidence for a Trace Species via Bayesian Model Comparison

Bayesian inference is a popular tool used to fit a model to data and is efficient in exploring the complex, multidimensional parameter space of models. Its use has been growing in analysis of spectroscopic/photometric data of solar system surfaces (e.g., J. Fernando et al. 2013; F. Schmidt & J. Fernando 2015; M. G. A. Lapotre et al. 2017; I. Belgacem et al. 2020, 2021; I. Mishra et al. 2021b, 2021a). At the heart of Bayesian inference is *Bayes's theorem*:

$$p(\theta|d, M) = \frac{p(d|\theta, M)p(\theta|M)}{\int p(d|\theta_M)p(\theta|M)d\theta} \equiv \frac{\mathcal{L}(d|\theta, M)\pi(\theta_j|M)}{\mathcal{Z}(d|M)}, \quad (3)$$

where $M$ is the model, $\theta$ is the set of free parameters of the model, $d$ are the data, $\mathcal{L}(d|\theta, M)$ is the likelihood probability function, $\pi$ is the prior, $\mathcal{Z}$ is the Bayesian evidence, and $p(\theta|d, M)$ is the posterior probability function. Since we assume the errors on our data to be Gaussian and independent, the likelihood function is also a Gaussian equal to

$$\mathcal{L}(d|\theta, M) = \prod_{k=1}^{N_{\rm obs}} \frac{1}{\sqrt{2\pi\sigma_k^2}} \exp\left(-\frac{\chi^2}{2}\right), \quad (4)$$

where $N_{\rm obs}$ is the number of observed data points (i.e., number of wavelength channels/data points in the observed spectrum), $\chi^2$ is the familiar goodness-of-fit metric, and $\sigma$ is the standard deviation of the Gaussian errors. For all the Bayesian analyses presented in this work, the free parameters $\theta$ are the abundances and average grain sizes of all the species in the model. We assume a uniform prior function $\pi$ for the free parameters (0–1 for all the abundance fraction parameters, and 10–1000 $\mu$m for all the grain size parameters). Since our prior function $\pi$ is uniform, the posterior distributions are entirely governed by the likelihood function $\mathcal{L}$, making our method equivalent to maximum likelihood estimation.

The numerator in Equation (3) determines the shape of the posterior probability function $p(\theta|d, M)$ and thus helps in parameter estimation. Consequently, the posterior distributions (see Section 3.2 for an example) of the free parameters in our model are also determined by this numerator. The denominator in Equation (3), called the Bayesian evidence $\mathcal{Z}$, while irrelevant for parameter estimation, is crucially useful in comparing models (e.g., I. Mishra et al. 2021b, 2021a). In BMC, we compare the probability of two models, $M_1$ and $M_2$, given the data. Using the Bayes rule, the ratio of these probabilities can be broken down in the following way:

$$\frac{p(M_1|d)}{p(M_2|d)} = \frac{p(d|M_1)}{p(d|M_2)} \frac{p(M_1)}{p(M_2)}$$
$$= \frac{p(d|M_1)}{p(d|M_2)} \times 1 = \frac{\mathcal{Z}(d|M_1)}{\mathcal{Z}(d|M_2)}. \quad (5)$$

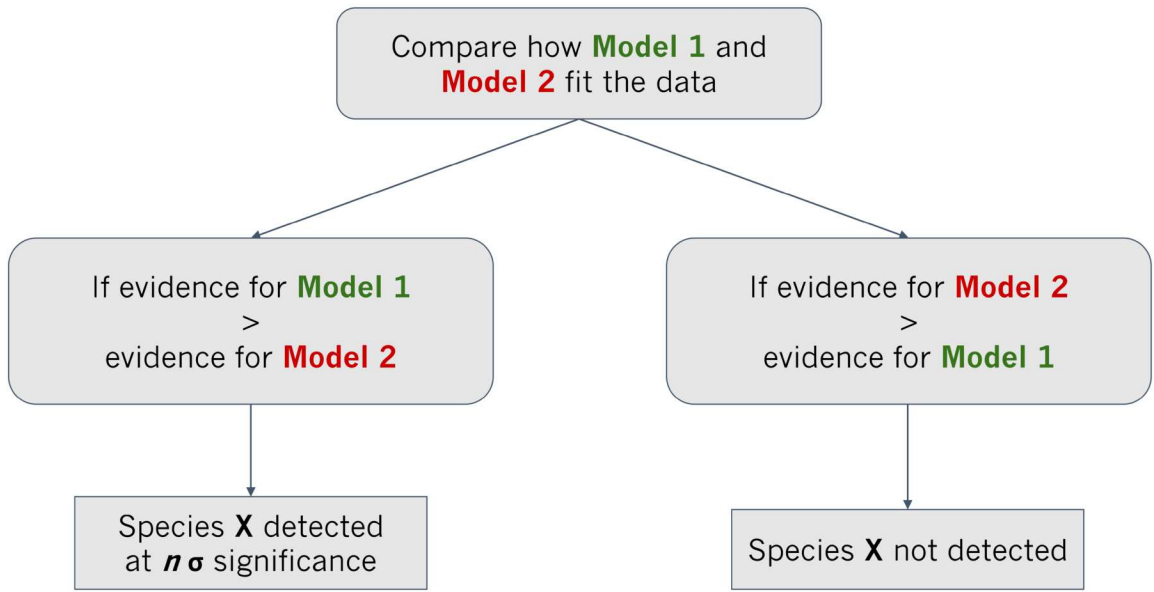

**Figure 3.** The process of determining evidence of a species in spectroscopic data using BMC. As described in Section 2.3, a Bayesian inference exercise allows us to calculate the Bayesian evidence $\mathcal{Z}$ for any model, i.e., the model's probability given the data. We compare the $\mathcal{Z}$ values of two models: Model 1 with all our candidate species, and Model 2 with our species of interest X excluded. If Model 1's $\mathcal{Z}$ is greater than Model 2's $\mathcal{Z}$, then we conclude that the presence of species X in Model 1 is helping it outperform Model 2, and hence there is evidence for species X at an $n\sigma$ significance. The value of $n$ will be higher if the difference in the $\mathcal{Z}$ values of the two models is higher. On the other hand, if Model 2's $\mathcal{Z}$ is greater than Model 1's $\mathcal{Z}$, then we conclude that the exclusion of X in Model 2 helped it outperform Model 1 and thus there is no evidence for species X.

In this equation, we assume that the prior probability of both models is the same ($p(M_1) = p(M_2)$), and in the final step we recognize that the likelihood of the data given the model, $p(d|M)$, is the same as the integral in the denominator of Equation (3), i.e., the Bayesian evidence $\mathcal{Z}$. Hence, in BMC, we compare two models by comparing their Bayesian evidence.

For our specific purpose of analyzing spectroscopic data, BMC comes in very handy to quantify our confidence in the presence/absence of any candidate species. The flowchart in Figure 3 describes this process. We compare two models: Model 1, which has all our candidate species, and Model 2, which excludes one candidate species X. If Model 1 turns out to have a greater value of $\mathcal{Z}$ as compared to Model 2, we take it as evidence for the presence of X in the data, whereas if it is vice versa, we conclude that there is no evidence for the presence of X in the data. In the case where the ratio $\mathcal{Z}_{\text{Model}-1}/\mathcal{Z}_{\text{Model}-2} > 1$, values in the range of 1–2.5, 2.5–12, 12–150, and >150 would be considered as "marginal," "weak," "moderate," and "strong" evidence for species X, respectively, as per the conventional ranges (H. Jeffreys 1998; R. Trotta 2008; B. Benneke & S. Seager 2013). This ratio, also known as the *Bayes factor* ($B_{12}$), can be converted to the familiar $\sigma$ significance metric of detection (T. Sellke et al. 2001; R. Trotta 2017), which we will henceforth refer to as $\sigma_{\text{bmc}}$ (where "bmc" stands for BMC), using

$$\mathcal{B}_{12} \leqslant -\frac{1}{ep\ln p} \quad (6)$$

$$p = 1 - \text{erf}\left(\frac{\sigma_{\text{bmc}}}{\sqrt{2}}\right), \quad (7)$$

where $p$ is the "$p$-value" and "erf" is the error function. Hence, Equation (6) converts the Bayes factor to an upper bound on $p$-value, which in turn gives a lower bound on the detection significance through Equation (7). It should be noted that Equation (6) is valid only for $p \leqslant e^{-1}$ or equivalently $\mathcal{B}_{12} \geqslant 1$. We refer the reader to Section 2.2.3 in I. Mishra et al. (2021a) for a table of Bayes factor values and their corresponding $\sigma_{\text{bmc}}$ values.

### 2.4. Evaluating the Limits of Detection of an Instrument via "Detection Significance Heatmaps"

While Figure 2 presents an example of a simulated reflectance spectrum of a trace species (HCN) in a mixture with water ice for a fixed abundance of HCN and fixed SNR of data, useful insights can be gained from studying how the spectroscopic features of a trace species change as we vary its abundance and the SNR of the data. These two parameters primarily control how well one is able to pull out the "signatures" of any species from a spectrum.

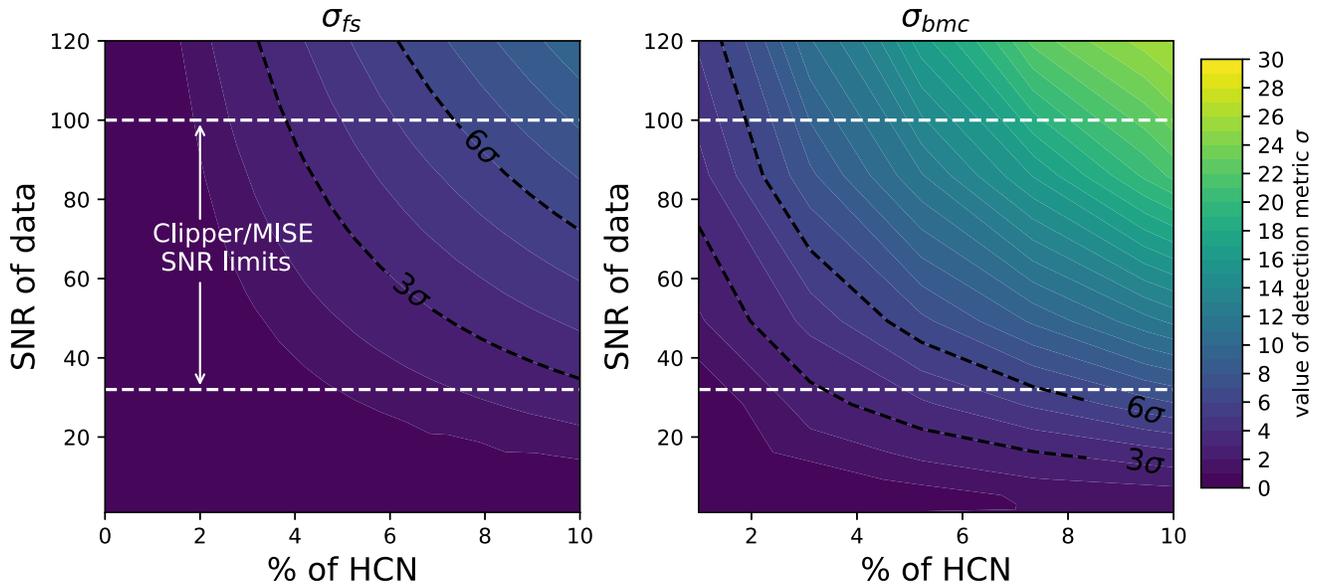

**Figure 4.** Left panel: a "detectability heatmap" of the $\sigma_{fs}$ metric of detection of HCN in a simulated spectrum of water and HCN mixture, derived from the average strength of the prominent absorption features of HCN (see Section 2.2). The abundance fraction of HCN is on the *x*-axis, and the SNR of the simulated data is on the *y*-axis. The white dashed lines mark the limits of expected SNR for Europa Clipper's MISE spectrometer in the 3–5 μm wavelength range. The $3\sigma$ and $6\sigma$ contours, conventional limits for tentative and strong detection, respectively, are shown with black dotted lines. Right: detectability heatmap of the $\sigma_{bmc}$ metric for detection of HCN in simulated data of water and HCN mixture, calculated using the BMC approach, as described in Section 2.3. The BMC-based detection pushes the $3\sigma$ and $6\sigma$ threshold to lower HCN abundances, closer to ~1%, which is typically the organic abundance on icy surfaces.

In Sections 2.2 and 2.3 we presented two metrics, $\sigma_{fs}$ and $\sigma_{bmc}$, that help us evaluate the evidence of a given species in reflectance spectra. The dependence of $\sigma_{fs}$ and $\sigma_{bmc}$ on SNR of the simulated data and the abundance of the trace species can be visualized with a two-dimensional "detectability heatmap" (see the example presented in Figure 4). An (*x*, *y*) point on the 2D heatmaps corresponds to the significance of detection ($\sigma_{fs}$ or $\sigma_{bmc}$) of HCN in a reflectance spectrum of water ice and HCN, with *x* being HCN's abundance fraction and *y* being the SNR of the data.

These detectability heatmaps allow us to explore detection thresholds across a range of SNR and abundance levels. Europa Clipper's MISE spectrometer's mission requirement for the 3–5 μm region is an SNR of 20, but laboratory measurements show that it can actually achieve a higher SNR of 100 in this low-reflectance regime (D. L. Blaney et al. 2023). Figure 4 highlights that the absorption features of few-percent-by-abundance HCN, via the $\sigma_{fs}$ metric, are detectable at >3σ confidence within MISE's SNR limits. While, as expected, the detection significance increases with higher abundance of HCN and greater SNR of the data in both heatmaps, using the $\sigma_{bmc}$ metric improves the detectability of HCN significantly. In Section 3.5, we compare these thresholds to the effective capabilities of current and upcoming instruments, using a dilution-based SNR scaling framework to account for differences in spatial and spectral resolution.

## 3. Results and Discussion

In the next two subsections (Sections 3.1 and 3.2) we discuss detectability heatmaps, like in Figure 4, for all of the selected trace organic species (Table 1) in two-component mixtures with water ice. In Section 3.3, we explore their detectability in a multicomponent mixture, with water and all the trace species mixed together. In Section 3.4, we discuss the sensitivity of trace organic features in a reflectance spectrum on physical parameters in the Hapke model, such as observation geometry angles, grain size, and porosity.

### 3.1. Detectability of Trace Species via Measuring the Prominence of Their Features

For the trace organic species of interest in this work, Table 1 lists their absorption features beyond 3 μm. Figure 5 shows reflectance models of water ice mixed with each trace species at 20% abundance (30% for $NH_3$ and $C_2H_2$, as their features are the weakest among this pool). A pure water-ice spectrum and a "difference" spectrum between the two-component mixture and pure water-ice models are also shown in each panel to highlight the spectroscopic behavior of each trace species in the 3–5 μm wavelength region. Among the trace species considered here, HCN, $C_2N_2$, $C_2H_4$, $CH_3OH$, and $C_3H_6O$ have at least two sharp features. We note that for the calculation of the $\sigma_{fs}$ metric (see Equation (2)) we chose absorption features that are sharp and deep (shown in bold in Table 1). However, we tried other choices and combinations of features, and the $\sigma_{fs}$ did not vary significantly.

In Figure 6, the $\sigma_{fs}$ heatmaps for all our trace organic species are shown. Within MISE's detectability limits, HCN, $C_2N_2$, $C_2H_4$, and $CH_3OH$ reach the 3σ detection threshold of their sharpest features at ~5% abundance and SNR of ~100 (predicted ß upper limit of MISE). This follows from the fact that these four species have at least two sharp and deep features. The lowest-significance heatmaps belong to $NH_3$ and $C_2H_2$, whose prominent features are shallow and broad. Interestingly, $C_2H_6$'s detectability is also quite low, despite its spectral richness as seen in Figure 5. This is because two of the three features used for calculating its $\sigma_{fs}$ metric, at 3.78 and 3.91 $\mu$m, weaken significantly at abundances explored in the heatmaps (0%–10%), whereas the abundance of trace species in the models in Figure 5 is set quite high at 20% for easier visual identification of their prominent spectral features.

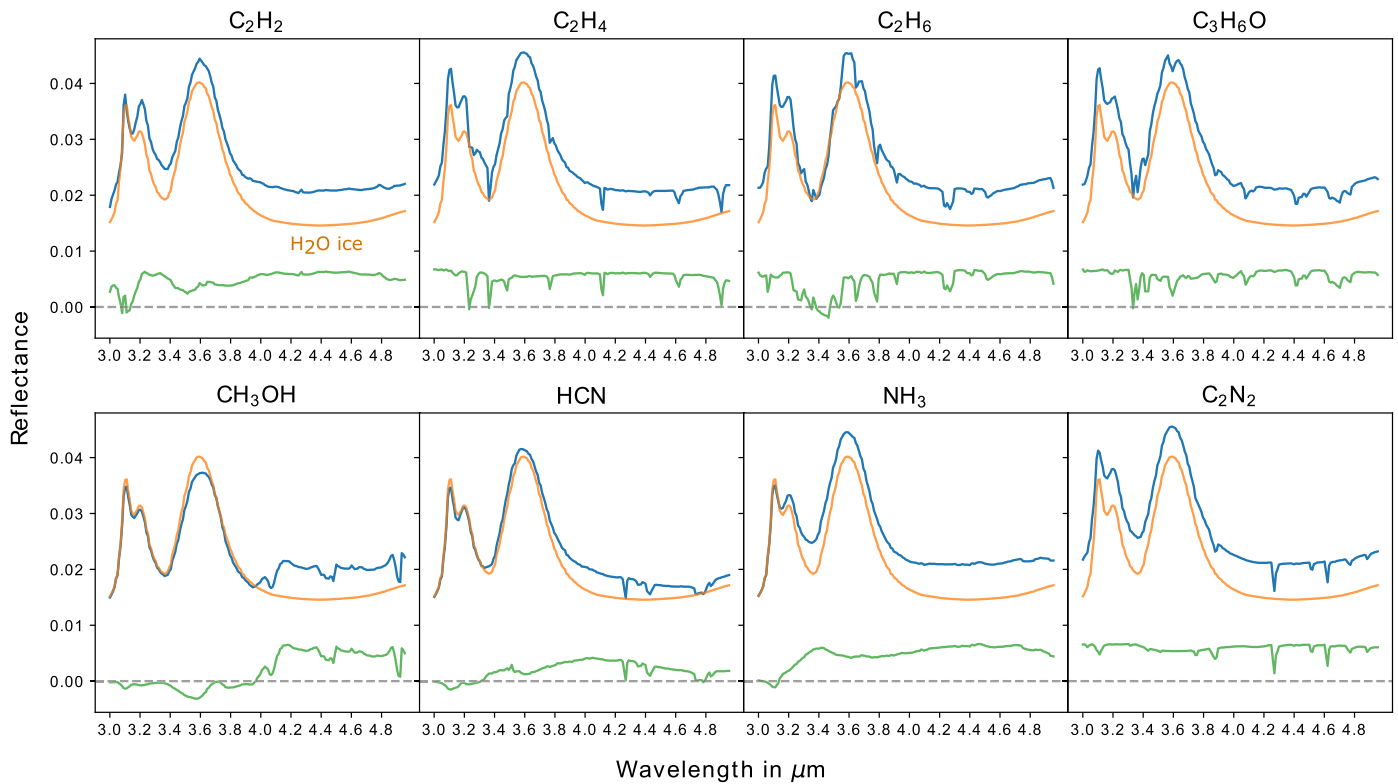

**Figure 5.** Reflectance spectrum models of two-component mixtures of water ice (80% abundance by number) and the labeled species (20% abundance by number) are shown in blue. The orange curve in each panel shows a pure water-ice spectrum for reference, while the green line is the residual, i.e., difference, between the blue and the orange models. The dashed gray line marks the zero reflectance level, as a reference for the residual curve. Qualitatively, large and "feature-rich" deviations of the residual curve from zero indicate a more dominant trace species, such as $C_2H_6$. The sharp features in the residual curve also highlight where the major absorption features of the trace species are present, since water's features are pretty broad. All models are at MISE's spectral resolution of 10 nm. Note: while the full 3–5 $\mu$m spectral range was used in all analyses, absorption features near 3.16 and 3.33 $\mu$m were not included in the $\sigma_{fs}$ calculations owing to overlap with broad water-ice bands (see Table 1).

### 3.2. Detectability via Bayesian Model Comparison Approach

An important insight from Figure 4 is that the BMC improves detectability of HCN to <1% abundance, especially within Clipper's SNR limits. Figure 7 presents detectability heatmaps of all the trace species considered here, evaluated using the $\sigma_{bmc}$ metric (Section 2.3). These heatmaps display higher detection significance values as compared to those presented in Figure 6. The detection threshold of 3$\sigma$ is achievable for a fraction-of-a-percent abundance of all the trace species, even toward the lower end of MISE's SNR limits. Although these detection limits correspond to a simple two-component mixture, the significantly high values of $\sigma_{bmc}$ (>6$\sigma$ for most part) indicate that BMC may still be able to reveal evidence of a species in a complex mixture of species with significant confidence. Detectability of individual trace species in a more complex mixture is discussed later in Section 3.3.

The BMC approach improves the detectability of a trace species significantly because it considers the effect of a species on the entire spectrum and not only at wavelengths where its sharp features are present. In the BMC exercise, we are testing for the hypothesis that a constituent species is present in the data by evaluating the change in the goodness of fit of the spectroscopic model with and without that constituent species. This includes not only the presence of absorption features at the correct wavelengths but also the absence of spurious features—i.e., candidates that introduce spectral features not seen in the data will be disfavored. If our species of interest is indeed present in the data, then its effect is apparent not only where its sharp features are present but also at other wavelengths.

An example of this effect is presented in Figure 8. Simulated reflectance data of a water and HCN mixture are fit with two models, one with both water ice and HCN present and the second with only water ice present. The residuals of the fit with the water-ice-only model show a systematic deviation from zero at all wavelengths, unlike the residuals of the fit with the water-ice + HCN model, which are randomly distributed around zero. The systematic effect in the former is due to the missing HCN, which has a noticeable effect on the continuum of the total reflectance spectrum despite its low abundance. BMC is able to pick out this tiny effect, along with the presence of the sharp absorption features at locations matching HCN's features, and quantifies this effect as a detection significance metric. Hence, the $\sigma_{bmc}$ detection metric for HCN is a significant improvement when compared to $\sigma_{fs}$ for a given abundance and SNR, as the latter solely relies on the data near the absorption features.

It is important to note that this is an idealized case, where the model components and the simulated data share the same photometric model and optical constants. In realistic applications, imperfect knowledge of spectral end-members—such as differences in temperature, phase, or composition—can produce model–data mismatches even when the correct components are included. However, these residuals can themselves be scientifically valuable, helping to identify missing components, inaccurate photometric assumptions, or the need for improved laboratory measurements of optical constants under Europa-like conditions.

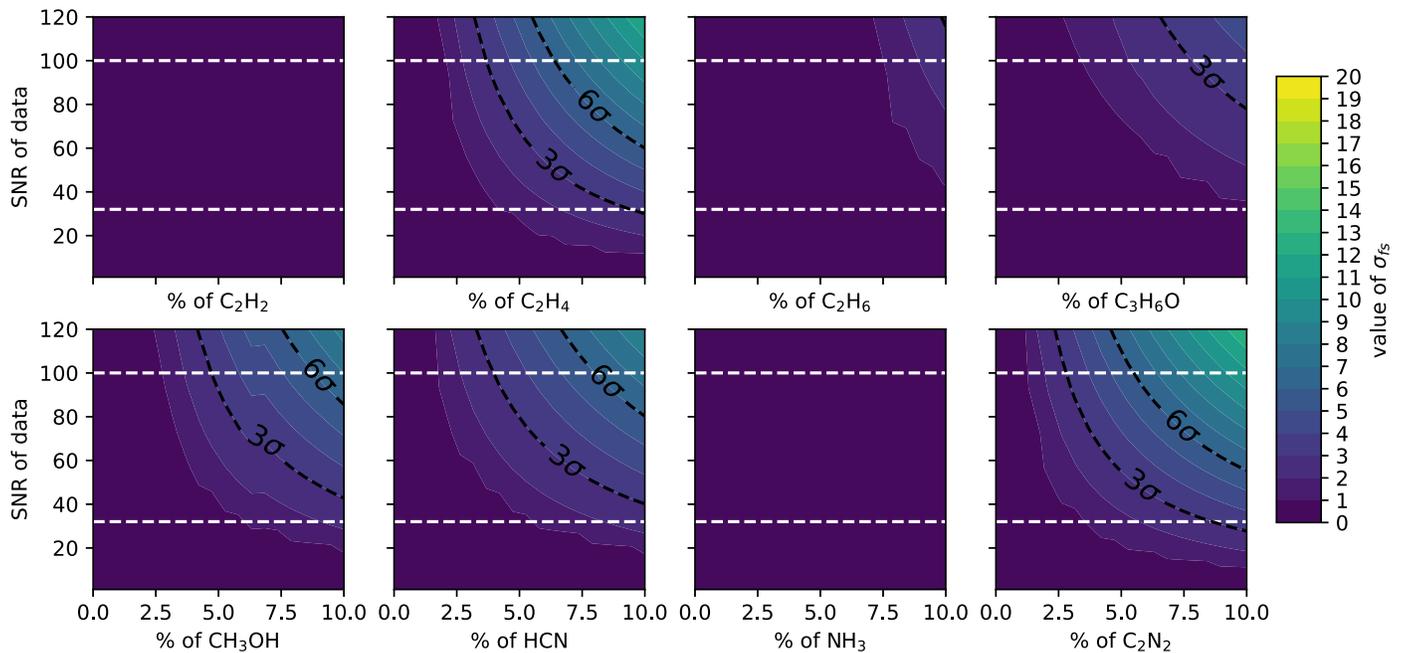

**Figure 6.** A collage of $\sigma_{fs}$ heatmaps, as in Figure 4, of various trace organic species in a two-component model with water ice. The spectroscopic features used for each species in the calculation of the $\sigma_{fs}$ metric are shown in Table 1. The white dashed lines mark the limits of expected SNR for MISE data in the 3–5 μm, and the black dashed lines mark the 3σ and 6σ contours.

Since BMC accounts for all wavelengths, we can infer perhaps why HCN's $\sigma_{bmc}$ heatmap in Figure 7 is the "weakest" of the lot. In Figure 5, when we look at the residual or the "difference" spectrum between the water-ice + HCN model and the pure water-ice model, we see that the magnitude of HCN's effect is milder as compared to other trace species. This could explain why the overall $\sigma_{bmc}$ heatmap for HCN looks weaker as compared to the other trace species. In contrast, while $NH_3$'s $\sigma_{fs}$ heatmap is pretty weak, on account of its two prominent spectroscopic features being very weak, it has a stronger $\sigma_{bmc}$ heatmap as compared to HCN. This could be because $NH_3$ is strongly affecting the continuum of the mixture spectrum below 4 μm, especially near the water absorption band at ~3.4 μm, as is evident from its residual spectrum.

### 3.3. An Example Case of Multiple Trace Species Mixed Together

In the previous sections, we investigated the detectability of trace organic species in a simple mixture with only water. The surface of Europa, however, is chemically complex (R. W. Carlson et al. 2009), and degeneracy between spectrally similar components, or components with overlapping spectral features, makes the inference of composition challenging (J. B. Dalton 2007). Here we consider a complex mixture—multiple trace species mixed together in a background of water ice—and investigate the detectability of each trace species with BMC. Along with our selected trace organic species (Table 1), we also add $CO_2$ and $SO_2$ into the mix.

We include $CO_2$ and $SO_2$, as both these oxides have features in the region beyond 4.0 μm (see Figure 9), where our trace organic species also have prominent features (Table 1). Hence, we can investigate a simple case of detectability of trace organics in the presence of other non-water-ice species that have similar sharp features in the 3–5 μm region. $CO_2$ and $SO_2$ are also astrobiologically significant oxidants and might play an important role in creating an environment on Europa of chemical disequilibrium in its subsurface ocean (e.g., C. F. Chyba & C. B. Phillips 2001; K. P. Hand et al. 2007). Moreover, $CO_2$ has also been linked to young geological regions on Europa like the chaos terrains (S. K. Trumbo et al. 2019), which indicates a possibly endogenic origin of the carbon that fuels a carbon cycle on Europa's surface, potentially resulting in trace organics of interest (R. W. Carlson et al. 2009; K. P. Hand & R. W. Carlson 2012). Published estimates of $CO_2$ abundances exist for the leading side of Europa, where the 4.25 μm feature in Galileo NIMS spectra (K. P. Hand et al. 2007) was used to constrain the abundance to ~360 ppm. For $SO_2$, a rough abundance estimate of ~0.1% comes from disk-integrated UV observations of Europa (A. R. Hendrix et al. 2011; T. M. Becker et al. 2022). Theoretical studies of $CO_2$ and $SO_2$ clathrate formation indicate that the bulk ice shell of Europa could have an oxidant concentration of up to ~7% (K. P. Hand et al. 2006).

Figure 10 presents a simulated reflectance spectrum of a mixture of 11 species—water ice, the eight trace organic species in Table 1, and the two oxides $CO_2$ and $SO_2$—at an SNR of 100 and spectral resolution of MISE (10 nm). All species except water are present at 1% abundance. Given the spectral complexity of trace organics in the 3–5 μm wavelength range (see Figure 1), we are more likely to have competing candidates for the same feature. For example, two absorption features in the data in Figure 10 could point to HCN, $C_2H_2$, $C_2H_4$, $C_2H_6$, $C_2N_2$, or $C_3H_6O$, all of which have overlapping features at those two wavelength locations. If we were looking at these data agnostically, we could come up with a pool of candidates for all absorption features and would want to systematically test the hypotheses for their presence in the data. Hence, this complicated spectrum provides a good test for our BMC approach, whose strength is to systematically evaluate the evidence of each species in the candidate pool.

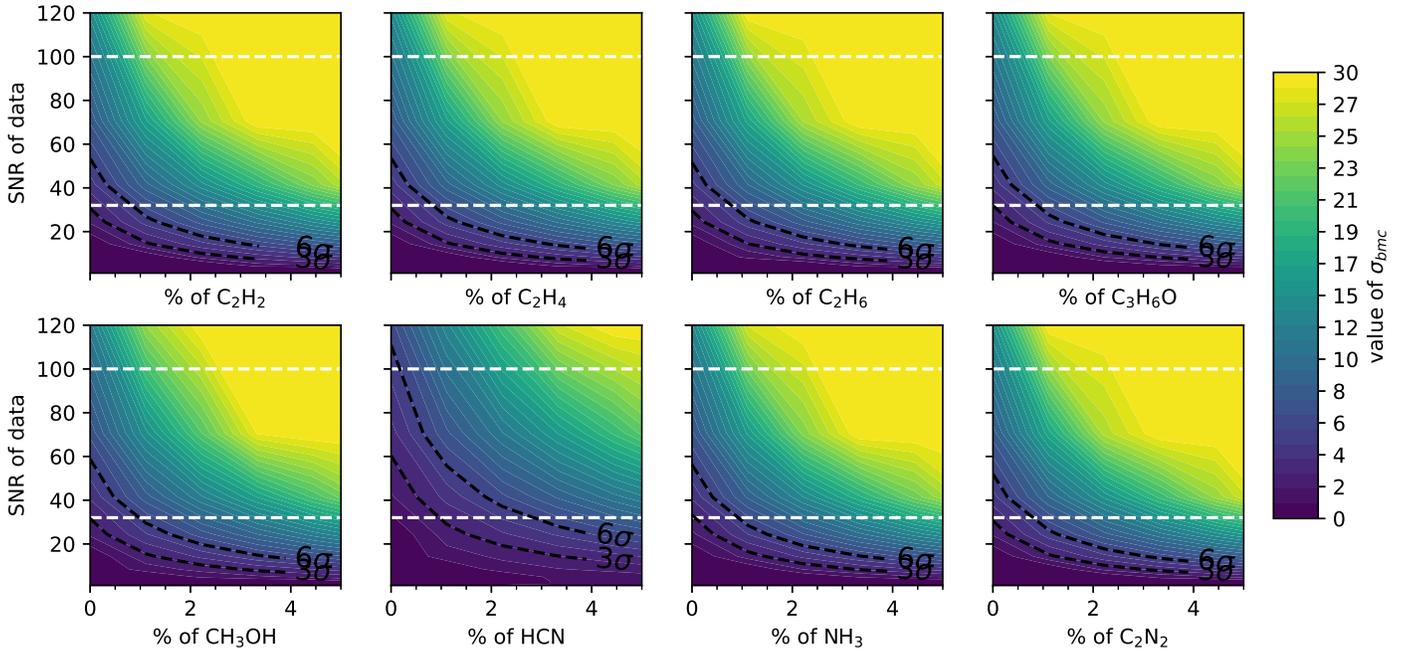

**Figure 7.** A collage of heatmaps for the $\sigma_{\rm bmc}$ metric for various trace organic species, in a two-component mixture with water ice, as in Figure 6, but now evaluated via the BMC approach (Section 3.2). All species have improved detectability as compared to the $\sigma_{\rm fs}$ metric of Figure 6. The latter only relies on the strength of the most prominent absorption features of a trace species, missing out on its effect at other wavelengths, which is picked up by the BMC approach.

The results of a BMC analysis of Figure 10's simulated data are presented in Figure 11. In this exercise, the reference model contains all 11 species, and we assume the abundances (10 parameters, as one abundance parameter is dependent) and grain sizes (11 parameters) as the free parameters, for a total of 21 free parameters. This reference model is compared to models that have one less species than the reference model, to allow us to evaluate the Bayesian evidence of each species as discussed in Section 2.3. The second column in the table in Figure 11 shows the detection significance ($\sigma_{\rm bmc}$) retrieved for all the trace species. They are all strongly detected in this mixture, with the exception of perhaps $NH_3$, which could be due to the fact that $NH_3$ lacks sharp features (see Figure 5). This is contrary to the much stronger detection significance of $NH_3$ we see in the heatmap in Figure 7, where a reflectance spectrum of SNR = 100 and $NH_3$ abundance of 1% shows $\sim 15\sigma$ evidence for $NH_3$. The stronger detection of $NH_3$ in the previous case is just due to the simplicity of the two-component spectrum. In the example presented in Figure 10, $NH_3$'s effect is much more diluted, and hence its retrieved detection significance is low. Of the two oxides, the retrieved detection significance of $SO_2$ is lower than that of $CO_2$. In Figure 9, we can see that $SO_2$'s effect on the water-ice + $SO_2$ spectrum is limited to its sharp features, whereas $CO_2$, along with having more features, strongly affects the continuum beyond 4.0 $\mu$m. We note that these results assume idealized conditions—accurate spectral end-members and perfect photometric modeling. As shown in Figure 12, small changes in photometric parameters such as grain size can significantly impact the shape and slope of the continuum. In this sense, our detection results should be viewed as best-case scenarios.

Just like in Figure 8, we can understand how a single species can have a detectable effect in our complicated mixture spectrum via Figure 13. It shows fits to the data with the reference model, which has all the species, and a model that does not have HCN. For the all-species model, the residuals are random, indicating a good fit. For the model without HCN, the fits show a systematic trend illustrating that HCN's absence affects the spectrum at many wavelengths. This systematic effect is efficiently picked up by BMC, resulting in high $\sigma_{\rm bmc}$ values.

Since our analysis shows moderate to strong evidence for all the candidates, we can correctly accept the 11-species model (water + 8 trace organics + $CO_2$ + $SO_2$) as our best hypothesis to explain the simulated data in Figure 10. Next, we can derive parameter estimations for our best model, and the top panel of Figure 11 presents the retrieved posteriors of abundances, in $\log_{10}$ space (the retrieved grain size distributions are not shown here). The median values, with the $\pm 2\sigma$ interval, of the abundance distributions are also shown in the third and fourth columns of the table in the bottom panel of Figure 11. The key observations from these posteriors are as follows:

1. The distributions for $C_2H_4$ and $C_2H_6$ are well constrained, and the true abundance value (1% for all) lies within the $\pm 2\sigma$ bounds, probably because these these species are among the most spectrally rich of all the trace organics we have selected for this exercise (see Figure 5). More spectral features of a species give the Bayesian algorithm more "levers" to constrain its abundance. They have strong features in the entire 3–5 $\mu$m wavelength region, as compared to the other species, whose features are mostly present beyond 4 $\mu$m.

2. Curiously, $CH_3OH$ also has a very well constrained and accurate abundance distribution. Unlike the hydrocarbons discussed above, it is spectroscopically weaker in the 3–4 $\mu$m region (Figure 5). On the other hand, $C_3H_6O$'s posterior is quite broad, despite its many spectral features across the entire 3–5 $\mu$m wavelength range. This contrast could be because $CH_3OH$'s effect on the continuum is pretty pronounced, as is evident from its residual spectrum in Figure 5, which has sharp curvatures in its continuum. Conversely, $C_3H_6O$'s residual spectrum has a flat continuum, which limits its effect to only its spectral features (which overlap with other species as shown in Figure 10) and could consequently weaken the constraint on its abundance.

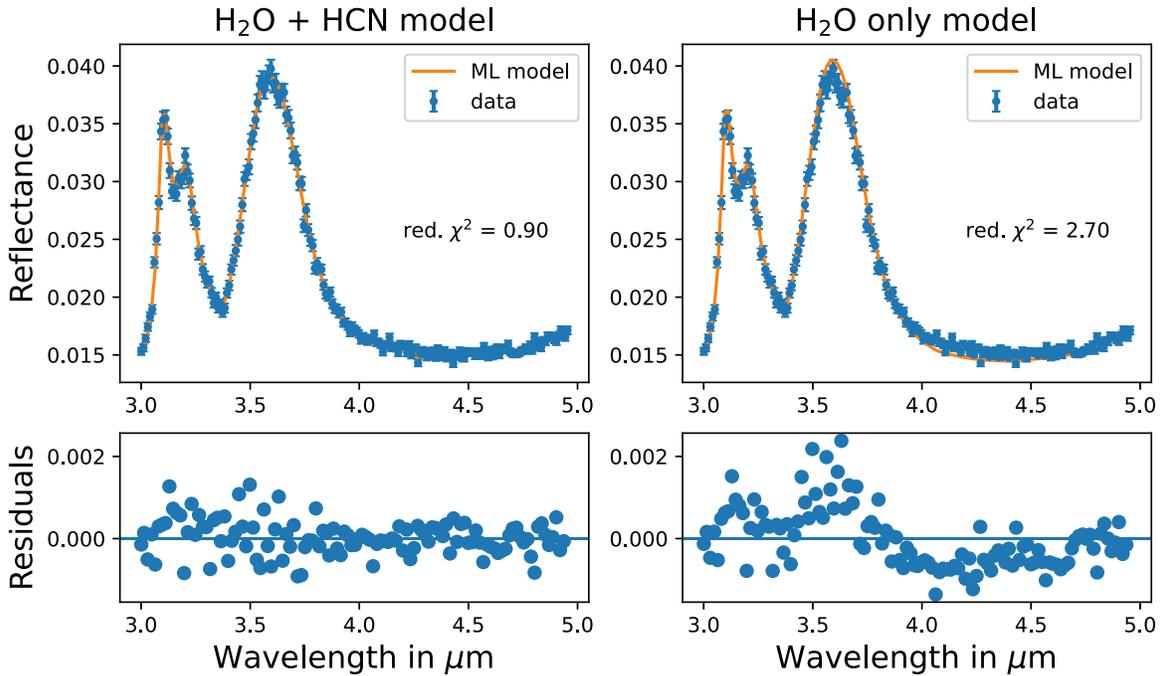

**Figure 8.** Best-fit or maximum likelihood solutions of a water+HCN model fit (left panel) and a water-only model fit (right panel) to simulated data (blue points). The simulated data are for a water and HCN mixture with 5% abundance fraction of HCN. Note that the Hapke model has been used to generate simulated data (see Figure 2), as well as for the model fits. The fit in the right panel is clearly inferior to the fit in the left panel, as illustrated by the residuals. Notably, the residuals in the right panel are poor and have a systematic trend caused by the missing HCN. The reduced $\chi^2$ values of the fits are also noted, which also indicate that the fit in the left panel is better. This example assumes perfect knowledge of the spectral end-members, which allows the correct model to fit the data extremely well. In real-world cases, imperfect end-members (e.g., wrong temperature or phase) may lead to residuals even in correct models, and such mismatches can provide valuable diagnostic information.

3. The two cyanides, HCN and $C_2N_2$, show contrasting distributions. While HCN's distribution is underestimated as compared to the true abundance value of 1%, $C_2N_2$'s distribution is overestimated as compared to the true abundance value. This contrast could be because HCN and $C_2N_2$ have an overlapping feature at 4.26 $\mu$m. HCN only has two strong features in the 3–5 $\mu$m region (at 4.26 and 4.42 $\mu$m), and the 4.26 $\mu$m feature is the stronger of the two. $C_2N_2$, on the other hand, is spectrally very rich (Figure 5), and its presence in the model is dominating the absorption in the data at 4.26 $\mu$m. Hence, HCN goes underestimated, and to compensate for that, $C_2N_2$'s abundance gets overestimated.

4. For the two oxides, $CO_2$ and $SO_2$, we can only obtain an upper limit on their abundances, rather than a well-constrained distribution. This is because both these species primarily affect the spectrum beyond 3 $\mu$m, with limited spectral features (Figure 9). The Bayesian algorithm thus has fewer "levers" to work with in order to get good constraints on their abundances.

### 3.4. Sensitivity of Detection to Observation Geometry and Other Model Parameters

In all the modeling analyses presented so far, we fixed key physical parameters of the Hapke equation (Equation (1))—the observation geometry angles (incidence, emergence, and phase), grain size, and porosity—and solely focused on the chemical parameters of interest, i.e., the abundances. It is important to consider the effects of these physical parameters on reflectance, as they can provide further insight into the detectability of trace species. Here we present a qualitative assessment of how these parameters may affect detectability, but we defer inclusion in the full BMC analyses to future work.

In Figure 12, reflectance models of a water and HCN mixture, with varying values of incidence angle, phase angle, grain size, and porosity, are shown. Emergence angle is not shown here, as its effect is similar to incidence angle. A few conclusions can be drawn from these sensitivity plots. First, we see that a smaller incidence (and emergence) angle leads to a more pronounced reflectance spectrum with deeper and sharper features. Hence, observations with less extreme incidence and phase angles are more desirable for ensuring that the trace species have pronounced features. Second, the phase angle and porosity both cause the reflectance spectrum to change its overall amplitude, independent of wavelength, but do not significantly affect the depth or sharpness of the HCN features. Finally, the grain size has a strong effect on the HCN features, with larger grain sizes leading to stronger and sharper features (closer to grain sizes of ~1000 $\mu$m; however, the band depths start decreasing again, as the overall reflectance goes down sharply).

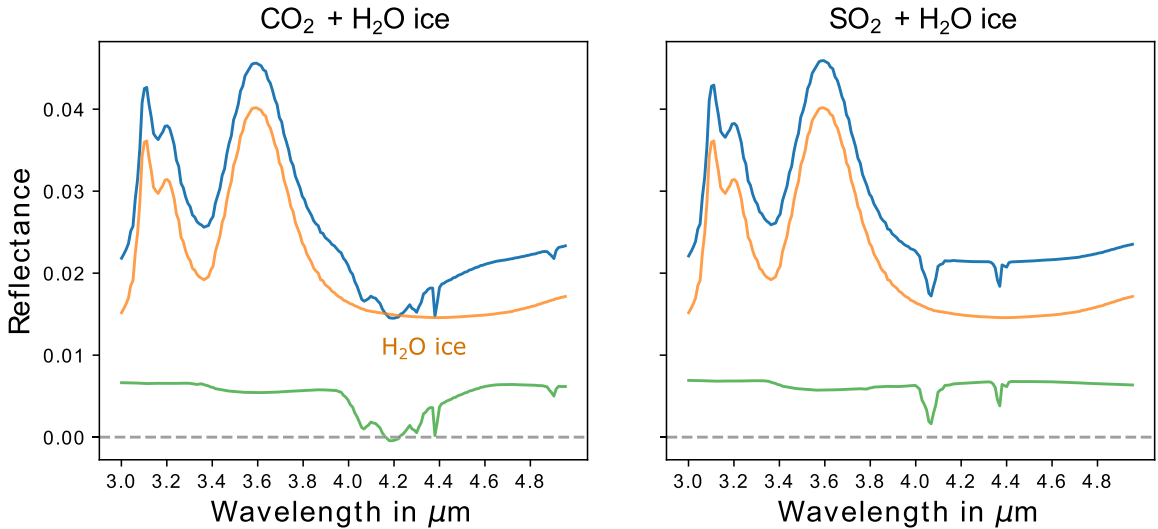

**Figure 9.** Simulated spectra of $CO_2$ (left) and $SO_2$ (right) mixed with water ice (blue) and their residuals (green) with a pure water-ice spectrum (orange), as in Figure 5. $CO_2$'s prominent features at 4.19, 4.38, 4.29, 4.90, and 4.06 μm and the $SO_2$'s prominent features at 4.06 and 4.36 μm stand out in the residual curves.

Interestingly, we find that increasing the grain size of the minor HCN component (while keeping its abundance fixed) can result in a higher total reflectance in the 3–5 μm region. This behavior arises from the interplay between two effects. First, although the single-scattering albedo $w$ of HCN *decreases* with increasing grain size owing to greater internal absorption, the Hapke model for intimate mixtures weights each component by the product of its number density and particle cross-sectional area, i.e., $f \cdot D^2$. Thus, larger HCN grains present more projected area and therefore contribute more strongly to the mixture's effective scattering behavior. Second, HCN has a higher single-scattering albedo than water ice in much of the 3–5 μm spectral region, for the whole range of grain sizes considered here (10–100 μm). Consequently, the growing contribution of HCN with grain size leads to an overall increase in reflectance in this region, despite HCN itself becoming more absorbing. This effect illustrates the importance of both optical constants and physical grain properties in determining detectability of trace components in intimate mixtures. Except for this sensitivity study, all the simulated data used in this work were generated using a grain size of 50 μm for both water and the trace component(s), which is on the lower end of ∼10−1000 μm that has been estimated for various parts of Europa's surface (J. B. Dalton et al. 2012; N. Ligier et al. 2016; O. Poch et al. 2018; I. Mishra et al. 2021b, 2021a). We chose 50 μm as the grain size in this work because the search for trace organics on Europa with Europa Clipper is primarily going to target low-irradiation regions, where organic signatures might be preserved (D. L. Blaney et al. 2017), and prior studies have suggested that the lack of particle bombardment may preserve smaller grain sizes (∼hundreds of microns) in these locations (R. N. Clark et al. 1983; T. A. Cassidy et al. 2013).

### 3.5. Dilution-based SNR Scaling and Instrument Comparisons

To interpret the detectability thresholds in our heatmaps within the context of current and planned observational facilities, we compare the effective ability of each instrument to detect highly localized features on Europa's surface. While many facilities achieve high SNRs, their larger spatial pixels and/or narrow wavelength bins can dilute spectral signals from small compositional anomalies.

Europa Clipper's MISE spectrometer is designed to achieve a mission requirement of SNR ∼ 20 in the 3–5 μm region, but laboratory measurements suggest that it can reach SNR ∼ 100 in low-reflectance conditions (D. L. Blaney et al. 2023). To fairly compare this to Galileo/NIMS, Keck/NIRSPEC, and JWST/NIRSpec, we apply a dilution-based scaling that reflects how much of a localized feature's signal is captured by each instrument:

$$\text{SNR}_{\text{scaled}} = \text{SNR}_{\text{measured}} \times \left(\frac{A_{\text{feature}}}{A_I}\right) \times \left(\frac{\Delta\lambda_{\text{feature}}}{\Delta\lambda_I}\right). \quad (8)$$

Here $\text{SNR}_{\text{measured}}$ is the instrument's reported or estimated SNR at its native resolution. $A_{\text{feature}}$ is the assumed area of the localized trace material (taken here as 100 km$^2$), and $A_I$ is the area of the instrument's spatial pixel. $\Delta\lambda_{\text{feature}}$ is the bandwidth of the absorption feature of interest (10 nm), and $\Delta\lambda_I$ is the spectral resolution (or channel width) of the instrument. We adopt values consistent with MISE's configuration: a 100 km$^2$ spatial area and a 10 nm bandwidth.

For Galileo/NIMS, we adopt an SNR of ∼5 based on observations from the E6-TERINC sequence analyzed in G. B. Hansen & T. B. McCord (2008). The spatial resolution was approximately 50 km (2500 km$^2$ pixel area), and the spectral resolution was ∼25 nm. For Keck/NIRSPEC, SNR values of 50–200 are reported depending on the integration time used during observations of Europa in the 3–4 μm region (S. K. Trumbo et al. 2017, 2019). The spectral resolution is $R \sim 2000$, equivalent to a channel width of about 2 nm, and the spatial resolution corresponds to a projected area of approximately 90,000 km$^2$. JWST/NIRSpec data are based on SNR estimates obtained using the publicly available JWST Exposure Time Calculator, specifically for GTO Program 1250 (G. Villanueva et al. 2017). For the G395H grating, which spans 2.87–5.27 μm, we adopt a spectral resolution of ∼1.5 nm and a representative SNR of 150. The spatial resolution of NIRSpec translates to a projected pixel area of approximately 122,500 km$^2$ on Europa's surface. These values are summarized in Table 2.

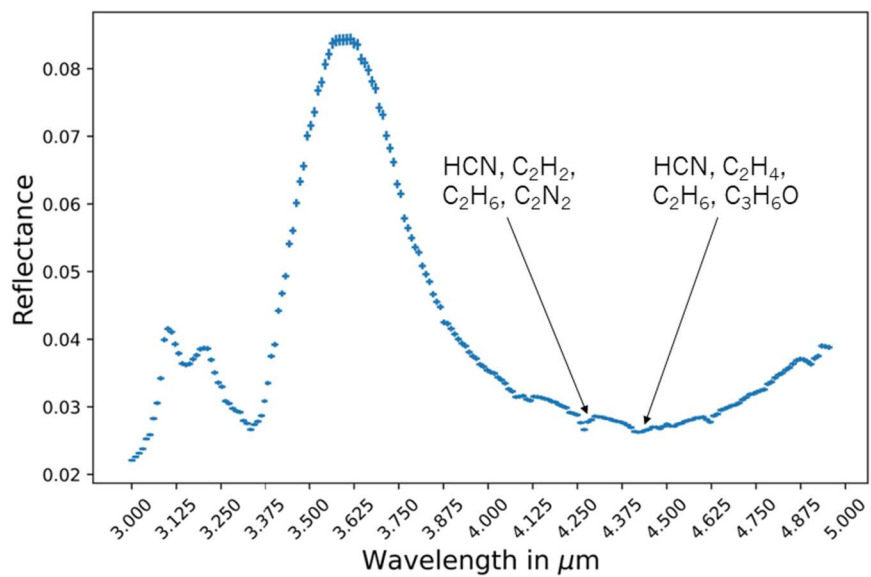

**Figure 10.** Simulated spectrum of water ice mixed with 8 trace organic species (Table 1), $CO_2$ and $SO_2$, at an SNR of 100 and at MISE's spectral resolution of 10 nm. Each of the 10 trace species has 1% abundance; hence, the mixture is 90% water ice and 10% non–water ice. Arrows point to two absorption features in the data near 4.25 and 4.42 μm, each with multiple possible trace species candidates. The 4.25 μm feature could also arise from $CO_2$, which has been detected on Europa. This example illustrates how the presence of overlapping features can lead to spectral confusion and underscores the importance of using statistical model comparison to evaluate candidate species.

**Table 2**
Comparison of Instrument Parameters and Their Scaled SNRs for Detecting Localized Trace Features on Europa

| Instrument | $SNR_{measured}$ | $A_I$ (km$^2$) | $\Delta\lambda_I$ (nm) | $SNR_{scaled}$ |
|---|---|---|---|---|
| Galileo/NIMS | 5 | 2500 | 25 | 0.08 |
| Keck/NIRSPEC | 50–200 | 90,000 | 2 | 0.28–1.11 |
| JWST/NIRSpec | 150 | 122,500 | 1.5 | 0.82 |
| MISE (Clipper) | 30–100 | 100 | 10 | 30–100 |

**Note.** The SNR values are scaled to match a hypothetical feature with spatial extent of 100 km$^2$ ($A_{feature}$) and spectral bandwidth of 10 nm $\Delta\lambda_I$, consistent with MISE's configuration. Please refer to Equation (3.5) for the definition of column parameters. The dilution-based scaling reflects how the detectability of trace organics diminishes when observed with instruments that have larger spatial pixels or finer spectral resolution. These values are meant to illustrate the importance of high-resolution measurements for detecting small-scale compositional heterogeneity.

The dilution-based detectability table highlights that, when a trace species is confined to a localized region (assumed here to be 100 km$^2$), instruments with large spatial and/or spectral resolution suffer significant dilution of the species' spectral signal. For example, even with a high per-pixel SNR, Keck/NIRSPEC's large spatial footprint causes its effective detectability to be significantly lower than MISE's. Similarly, even JWST/NIRSpec, despite its high SNR and spectral resolution, observes Europa with spatial pixels much larger than 10 km$^2$, leading to substantial dilution.

All scaled SNR values in the table are adjusted to match a feature of size 100 km$^2$ and bandwidth of 10 nm, consistent with MISE's configuration. This scaling assumes that the material is highly localized, occupying just one MISE-resolution element. Lower scaled SNRs in Earth- or space-based instruments highlight the severe spectral dilution when trying to detect trace organics on Europa using large pixels and narrow bins.

In reality, organic deposits on Europa are likely confined to even smaller areas, particularly along narrow lineae or banded terrain. Studies have shown that Europan lineae often have widths of 1–2 km, and in some cases even less (L. M. Prockter et al. 2017). Therefore, our assumption of a 100 km$^2$ deposit is likely conservative, and actual dilution effects could be much stronger than what is shown here. This further underscores the importance of high spatial resolution and motivates future efforts to isolate and target small-scale compositional anomalies using data from instruments like MISE.

In summary, this dilution-based SNR comparison demonstrates that Europa Clipper's MISE is uniquely capable of detecting localized organic signatures. While other observatories may achieve high raw SNRs, their spatial and spectral dilution severely hampers detectability for localized trace features. This further highlights the need for high-resolution spectroscopy at high spatial resolution to isolate trace species in regions like chaos terrain and linea, where endogenous material is likely to be exposed.

## 4. Conclusions

We have presented a study of detectability of trace organics in a background of water ice via NIR spectroscopy, specifically focused on Europa's surface and the capabilities of the upcoming Europa Clipper mission. The trace species we selected for this work, shown in Table 1, span different chemical bonds of interest found in organics, such as C–H, C=C, C≡C, C=O, and C≡N. We have presented two approaches for inferring the presence/absence of a trace species in a reflectance spectrum: (1) by calculating the average prominence of its absorption features ($\sigma_{fs}$ metric in Section 2.2), and (2) by comparing the ability of models with and without our species of interest to fit the data, via BMC ($\sigma_{bmc}$ metric in Section 2.3). Next, we applied these detection significance metrics to varying SNR of data and abundance of trace species, which can be visualized in the form of "detectability heatmaps" (Sections 3.1 and 3.2). SNR limits (at fixed spatial and spectral resolution) of various spectroscopic instruments can be plotted on top of these heatmaps to evaluate their ability to detect trace organics (e.g., the minimum abundance that can be detected at $3\sigma$ confidence). The detection significance heatmaps can also be used to compare different methods of detection, such as the two ($\sigma_{\mathrm{fs}}$ and $\sigma_{\mathrm{bmc}}$) discussed here. In Section 3.3, we discussed a complex example with many trace species mixed together with water ice and presented results for a BMC analysis of the complex spectrum. Finally, in Section 3.4, we examined the sensitivity of sharp organic features to physical parameters like observation geometry angles, grain size of regolith, and porosity of regolith.

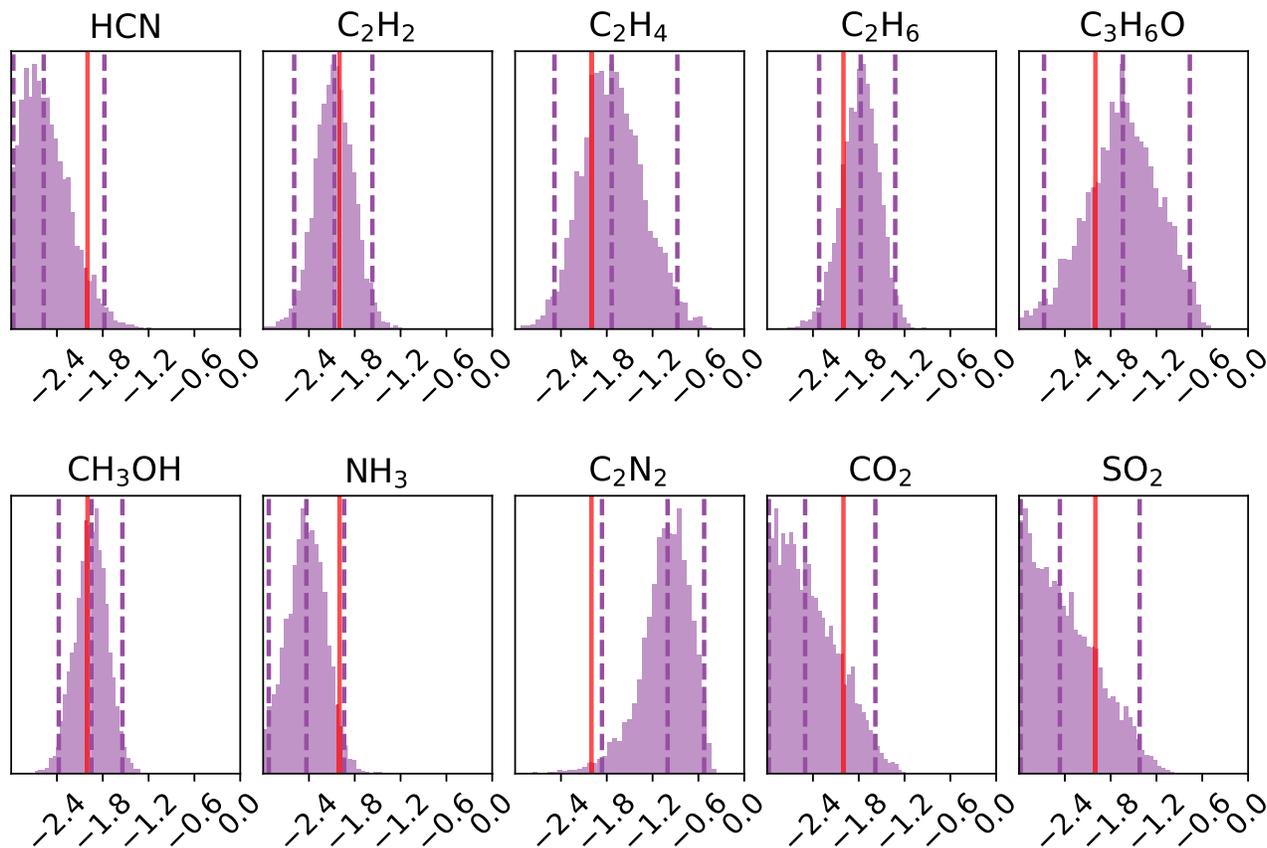

**Figure 11.** Results of a BMC analysis of the data in Figure 10 are presented here. Top: retrieved posterior distributions of the abundance of the trace species (in $\log_{10}$ scale). The dashed purple lines are the median values along with the $2\sigma$ upper and lower bounds (95% confidence intervals). The solid vertical red lines indicate the true abundance of the trace species (0.01 for all) used as inputs to generate the data in Figure 10. Bottom: the columns from left to right show the detection significance ($\sigma_{\mathrm{bmc}}$) and retrieved logarithmic and regular abundance estimates (median $\pm 2\sigma$) of each trace species.

The key conclusions from this work are as follows:

1. Detectability investigated with the $\sigma_{\mathrm{fs}}$ metric (Section 3.1), which depends on the strength of the absorption feature(s) of a species, shows that sharp and deep features, such as those of HCN, $C_2N_2$, $C_2H_4$, and $CH_3OH$, are detected more strongly as compared to species whose prominent features are shallow and broad, such as $NH_3$ and $C_2H_2$. Within Clipper/MISE's detectability limits, HCN, $C_2N_2$, $C_2H_4$, and $CH_3OH$ reach the $3\sigma$ detection threshold at ~5% abundance and SNR of ~100 (upper limit of MISE).

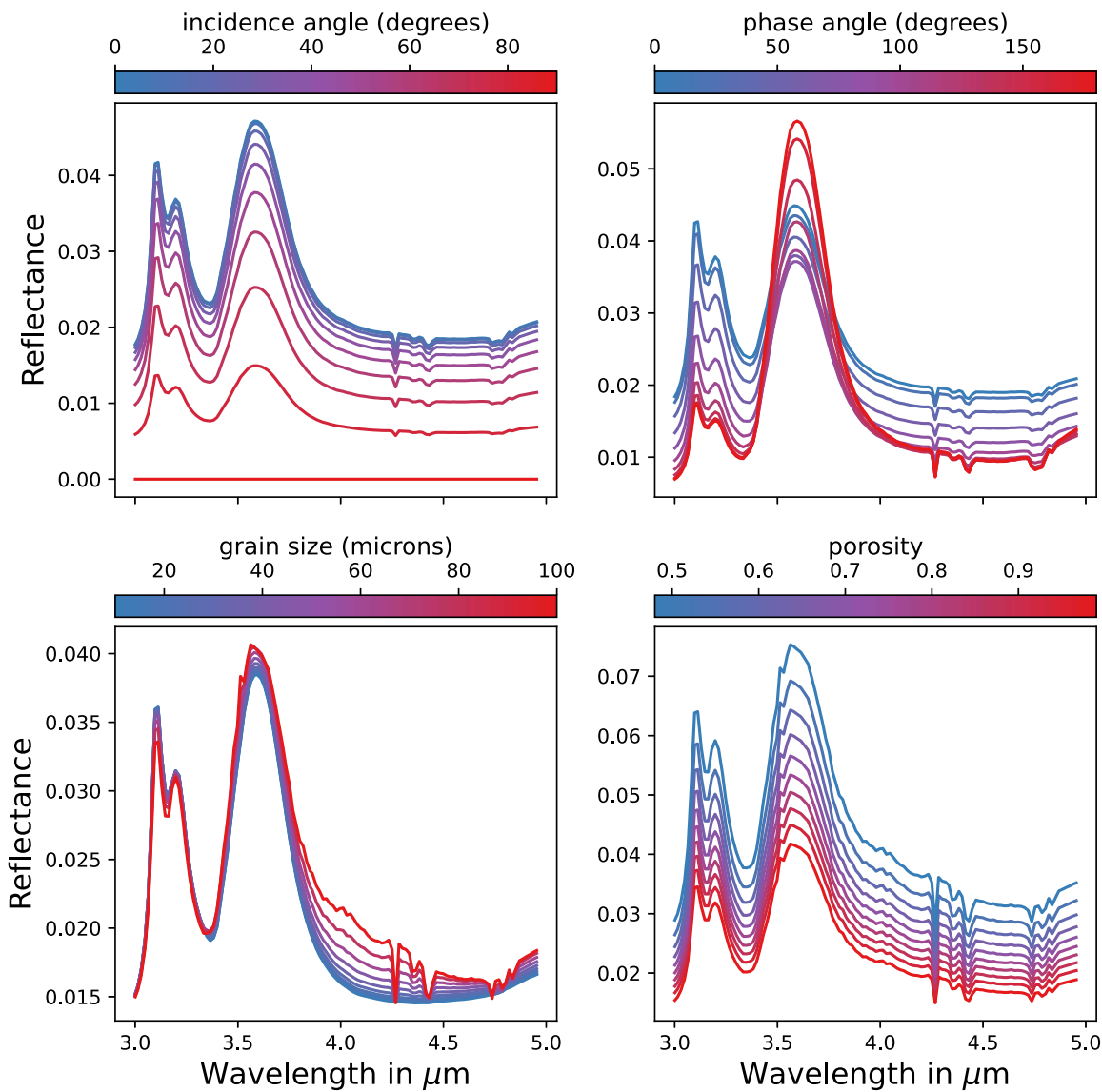

**Figure 12.** Sensitivity of the water + HCN spectrum to physical parameters in our model. From the top left and clockwise, a set of spectra for varying values of incidence angle, phase angle, porosity, and grain size (of trace species) are shown. Emergence angle sensitivity has not been shown here since it is equivalent to the incidence angle sensitivity. Phase angle and porosity have the effect of changing the overall amplitude of the spectrum. Incidence (and emergence) angle changes the amplitude of the features, with smaller values (less extreme geometry) more desirable for prominent features. Larger grain size of the trace species also leads to deeper and prominent features.

2. A Bayesian-evidence-based detection using the $\sigma_{bmc}$ metric improves the detection of all the trace organic species considered here (Section 3.2), with the detection threshold of $3\sigma$ achievable for a fraction-of-a-percent abundance, even toward the lower end of MISE's SNR limits (~30). The BMC approach improves the detectability of trace species significantly because it can quantify the presence of a candidate species in a spectrum by comparing fits of models with and without that species. In doing so, the effect of the species on the entire spectrum is evaluated, and not only at wavelengths where its sharp features are present.

3. A BMC approach is especially useful when considering a complex spectrum with many candidates, some of whom can have overlapping features. In that case, BMC helps us test the hypothesis for the presence of each candidate species in a systematic way. In the example presented in Section 3.3, we analyzed a simulated MISE reflectance spectrum of a mixture of 90% water ice and 1% each of 10 trace species with features in the 3–5 $\mu$m wavelength region. Our BMC analysis is able to detect all of the organic species at moderate to very strong detection significance ($\sigma_{bmc}$). The retrieved posterior distributions for the abundances of the trace species are also promising and show that we can go beyond detection to actually constraining abundances (to within $2\sigma$ of the true value) with MISE quality data. The most spectrally rich trace species, like $C_2H_2$, $C_2H_4$ $C_2H_6$, $CH_3OH$, and

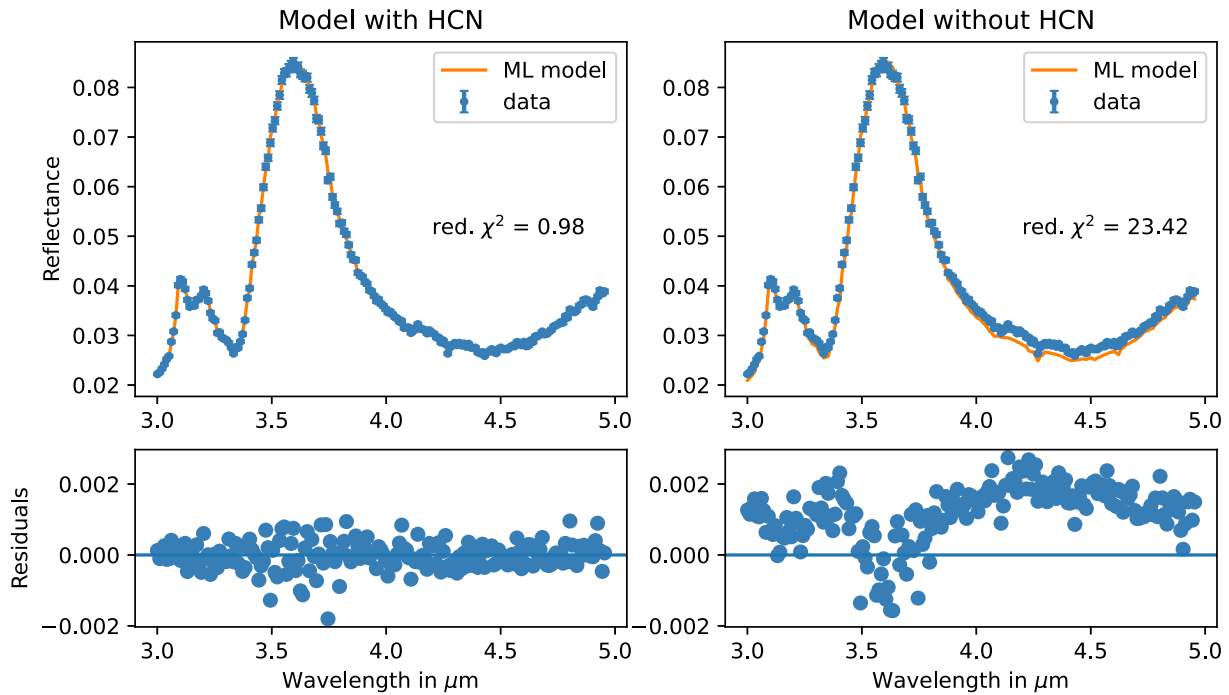

**Figure 13.** Maximum likelihood (ML) solutions for two models and the residuals of their fit to the simulated data from Figure 10 are shown. Note that the Hapke model has been used to generate simulated data (see Figure 2), as well as for the model fits. The model on the left has all 10 trace species (and water) from Table 1. The model on the right has all of them as well, except HCN. While the random distribution of the residuals on the left signifies a good fit, the residuals on the right clearly show systematic trends caused by the removal of HCN. This signifies the effect of HCN on the continuum of the total spectrum, just like in Figure 8, resulting in a strong detection of HCN ($>8\sigma$) when the two models shown here are compared. The reduced $\chi^2$ values of the fits also show how the fit suffers by excluding HCN from the model.

$C_2N_2$, have sharp and accurate posteriors, whereas for spectroscopically weaker species like HCN, $CO_2$, and $SO_2$, the posteriors provide an upper limit to their abundance.

4. These results for the detectability threshold of Clipper/MISE are encouraging because of the high spatial resolution of MISE's observations (<10 km pixel$^{-1}$ for global mapping), which will allow correlations to be drawn between the composition, and geological features like linea and chaos terrains, where endogenic material from the subsurface ocean is likely present.

5. While the focus of this work is on the abundance of trace species and how that affects the reflectance spectrum, physical parameters like observation geometry angles, grain size of regolith, and porosity of regolith, which were fixed to moderate values in this analysis, also affect reflectance spectrum. A sensitivity study (Section 3.4) shows that observation geometry angles, primarily the incidence and emergence angles, affect the sharp organic absorption features. Observations taken at moderate values of incidence and emergence angles, close to 45°, result in the most prominent spectral features and higher overall reflectance. Phase angle and porosity increase or decrease the overall reflectance in the 3–5 $\mu$m region but do not affect the individual spectral features. Finally, we find that larger grain sizes produce stronger absorption features. While the estimated grain size of Europa's regolith has a wide range (tens to hundreds of microns), even at a moderate grain size of 50 $\mu$m used in this work the detectability results for trace organic species are promising. We note, however, that this trend does not imply an overall brightening with increasing grain size—rather, the apparent increase in feature strength in our simulations arises from the higher single-scattering albedo of some trace species (such as HCN) relative to water ice in the 3–5 $\mu$m region and their increased contribution to the reflectance in the mixture due to the Hapke model's size-dependent weighting.

### 4.1. Future Work

A natural next step for the work presented here would be to consider more species in the mixture, such as other major Europan background species (e.g., hydrated sulfates, hydrated chlorinates, hydrated sulfuric acid). However, a fundamental bottleneck in numerically modeling reflectance spectra is the limited availability of optical constants of Europa-relevant species, measured in the laboratory at relevant temperature and pressure conditions (~100 K and vacuum). The effect of the temperature dependence of optical constants on our analysis, while minor (J. B. Dalton 2010), still needs to be explored. Future work that expands on either the species or wavelengths consider here would be best served by additional laboratory-derived optical constants obtained at Europan temperatures. Inclusion of additional species in our model would also make the Bayesian analysis more computationally expensive. With the 11 species model used in the analysis in Section 3.3, our BMC exercise took ~6 hr to run on 20 cores. The runtime of the Bayesian code we are using, `dynesty`,[6] scales in a sublinear way with the number of parallel threads (J. S. Speagle 2020). Hence, our code will need to be adapted for execution on a supercluster or GPUs.

---
[6] https://dynesty.readthedocs.io/en/stable/

Most importantly, we plan to test our analysis framework on laboratory reflectance data of Europa-like regolith, especially on samples that are irradiated with energetic particles. More broadly, further laboratory investigations (e.g., K. P. Hand & R. W. Carlson 2012) are needed to build and improve models that can connect the composition inferred from spectroscopy of Europa's surface to the composition of pristine material before irradiation. In addition, laboratory measurements of the optical constants of volatiles such as $CO_2$ and $SO_2$ in a range of physical phases (e.g., adsorbed on or trapped within water ice) are essential. Such data would allow future analyses to account for possible phase-dependent variations in spectral features, especially given the complex and possibly noncrystalline environments in which these volatiles may be found on Europa (G. L. Villanueva et al. 2023).

Our work here provides a useful and foundational insight into the capabilities of Europa Clipper's MISE instrument to improve and further constrain the composition of Europa's surface. As we show here, Clipper holds immense promise to open a new window in our understanding of trace species on Europa's surface. Along with Clipper, ESA's JUICE will also be present in the Jupiter system simultaneously and will carry out limited observations of Europa (O. Grasset et al. 2013). Although considering the exact science yield for the IR observations of Europa that will be provided by JUICE's MAJIS spectrometer (F. Poulet et al. 2023) is beyond the scope of the current work, our findings and methods are extensible to a similar analysis for MAJIS. The tools we present here can help us get the maximum information out of the data that Clipper and JUICE will provide, and they go beyond detection to constraining the abundances of more species on Europa's surface. Moreover, this analysis can be easily extended to other instruments and planetary bodies and also be used to set instrument parameter requirements (such as SNR) for future exploration of planetary surfaces in the solar system.

## Acknowledgments

I.M. was supported by the National Aeronautics and Space Administration (NASA) FINESST grant 80NSSC20K1381 for part of the work. Most of the manuscript preparation and revisions were carried out at the Jet Propulsion Laboratory, California Institute of Technology, under a contract with the National Aeronautics and Space Administration (80NM0018D0004). J.L. acknowledges the funding provided by Europa Clipper MISE project through the Jet Propulsion Laboratory, California Institute of Technology subcontract 1532536. K.P.H. acknowledges support from the Europa Clipper SUDA instrument and from the Jet Propulsion Laboratory, California Institute of Technology, under a contract with the National Aeronautics and Space Administration (80NM0018D0004). I.M. also thanks Geronimo Villanueva for help with understanding JWST's GTO observations of Europa, Samantha Trumbo for valuable discussions regarding Keck's observation parameters, and finally Ryan MacDonald for discussions regarding Bayesian statistics.

*Repository:* Cosmic Ice Laboratory (https://science.gsfc.nasa.gov/691/cosmicice/constants.html), https://www.sshade.eu/ (B. Schmitt et al. 2018).

*Software:* NumPy (C. R. Harris et al. 2020), Jupyter (T. Kluyver et al. 2016), Matplotlib (J. D. Hunter 2007), SciPy (P. Virtanen et al. 2020), dynesty (J. S. Speagle 2020), corner (D. Foreman-Mackey 2016), spectres (A. C. Carnall 2017).

## ORCID iDs

Ishan Mishra  https://orcid.org/0000-0001-6092-7674
Nikole Lewis  https://orcid.org/0000-0002-8507-1304
Jonathan Lunine  https://orcid.org/0000-0003-2279-4131
Kevin P. Hand  https://orcid.org/0000-0002-3225-9426